\documentclass[aps,twocolumn,superscriptaddress,prb]{revtex4-1}
\usepackage{color} 
\usepackage[dvipdfmx]{graphicx}
\usepackage{dcolumn} 
\usepackage{bm} 
\usepackage{amsmath,amsthm,amssymb}

\begin{document}

\title{
Reservoir computing by thin film embedded with magnetic impurities
}

\author{Shuto Kamakura}
\affiliation{%
Department of Physics, Toho University, 2-2-1 Miyama, Funabashi 274-8510, Japan
}

\author{Tomi Ohtsuki}
\affiliation{%
Physics Division, Sophia University, Tokyo 102-8554, Japan
}

\author{Jun-ichiro Ohe}
\email[junichirou.ohe@sci.toho-u.ac.jp]{}
\affiliation{%
Department of Physics, Toho University, 2-2-1 Miyama, Funabashi 274-8510, Japan
}


\date{\today}

\begin{abstract}
The reservoir computing based on the thin film embedded with magnetic impurities in the presence of the long-range (the dipole-dipole) interaction is numerically investigated.
We simulated the magnetization dynamics by taking into account the dipole-dipole interaction and performed the handwritten-digit recognition task.
Although the training data is prepared by taking spatial average in the sample, the high classification accuracy is achieved.
Our result demonstrates that the long range interaction effectively encodes the complex spatial input pattern into the time domain, even when only a spatially averaged output is accessible.
The proposed system paves the way for easily realizable magnetic reservoir computing.
\end{abstract}


\maketitle

Reservoir computing (RC) is a neuromorphic computing paradigm that utilizes a physical dynamic system, called a reservoir, to process time-varying inputs \cite{Jaeger2001, Maass2002}.
A key advantage of RC is that only the linear output layer requires training, while the complex system giving the non-linear dynamics of the reservoir are fixed\cite{Tanaka2019}.
This simplifies the learning process, drastically reduces computational resources and provides the low-power computing applications.
Recent studies have explored various physical systems as hardware reservoirs, including photonic integrated circuits, biological neural networks, and spintronic devices \cite{Torrejon2017, Yokouchi2023, Nakane2018, Nomura2019,Edwards2023,Allwood2023,Vidamour2023,Lee2022,Lee2023,Das2025,Tsunegi2019,Jiang2019,Hon2021}.
Magnetic systems are particularly promising due to their inherent nonlinearity, non-volatile nature, and scalability to the nanoscale.
Two examples of magnetic RC are systems based on (1) Spin-Torque Oscillators (STOs) and (2) Skyrmion Lattices.
The STOs shows non-linear microwave oscillations, providing a naturally rich dynamic response that has been successfully applied to tasks like speech recognition \cite{Torrejon2017,Tsunegi2019}.
However, the main challenges for STO-based RC include the high power consumption required to maintain continuous self-oscillation and the complexity of integrating multiple, coupled STOs for larger-scale systems.
The magnetic Skyrmion, which are topologically protected spin textures, offers an field induced complex dynamics.
Skyrmion-based reservoirs have demonstrated high accuracy in tasks such as handwritten digit recognition \cite{Yokouchi2023,Lee2022,Lee2023}.
To realize high-density and controllable skyrmion lattices at room temperature, however, remains a significant technological hurdle.

Here, we propose and numerically investigate a physical RC system that utilizes the collective magnetization dynamics of an array of interacting magnetic impurities.
There exists the dipole-dipole interactions, which are long-ranged ones, between magnetic impurities.
We simulate the magnetization dynamics by applying the spatial modulated magnetic field representing the input data.
The output data is the time dependent magnetization component that is averaged in the sample regions.
As in the case of ordinary RCs, we train the weight vector that connects the output data with the classification nodes, and obtained high accuracy for classifying the hand-written number.
We show that the dipole-dipole interaction makes the magnetization system suitable for performing RC.
Furthermore, by considering discrete magnetic impurities, the system naturally avoids the issue of exciting only a single magnon mode in continuous magnetic films.
The proposed system is easily fabricated and shows the high-accuracy results, and gives a significant step towards realizing practical machine learning by the spintronics device.
\begin{figure}[b]
\includegraphics[scale=0.35]{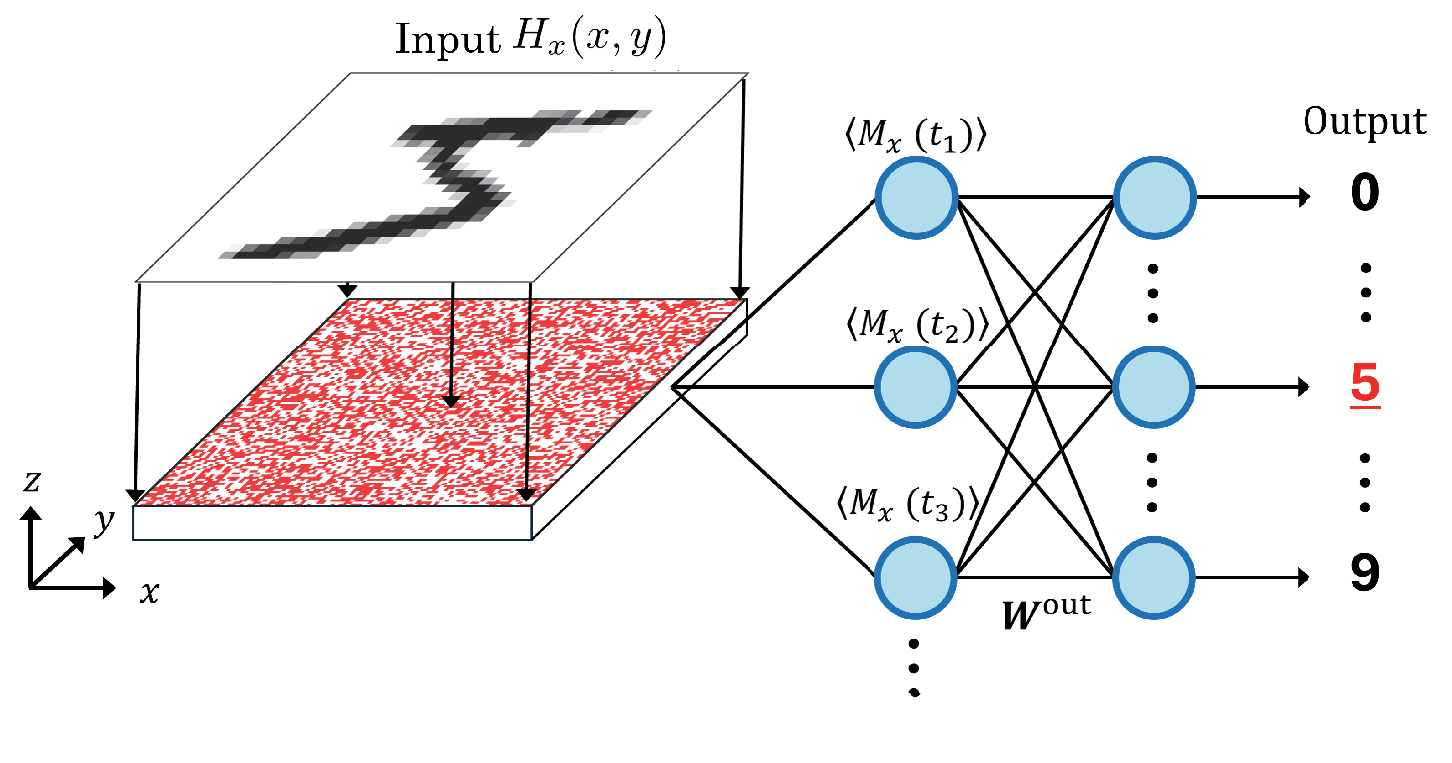}
\caption{
(color) Schematic of the thin film with magnetic impurities represented by red dots.
The applied magnetic field has a spatial dependence of the input hand-writing number.
The gray scale data of the image is converted into the $x$ component of the external field.
}
\label{fig1}
\end{figure}


The schematic of the proposed system is shown in Fig.~\ref{fig1}.
We use the thin film of the size $1$\,$\mu$m$\times 1$\,$\mu$m$\times0.1$\,$\mu$m with magnetic impurities.
The size of each impurity is $10$\,nm$\times 10$\,nm$\times 0.1$\,$\mu$m, and randomly distributed in the film.
The dynamics of magnetization are described by the Landau-Lifshitz-Gilbert equation,
\begin{equation}
\frac{\partial \mathbf{M}}{\partial t} = -\gamma \mathbf{M} \times \mathbf{H}_{\rm eff} + \frac{\alpha}{M_{\rm s}} \left( \mathbf{M} \times \frac{\partial \mathbf{M}}{\partial t} \right)
\end{equation}
where  
$\mathbf{M}$ is the magnetization vector, $\gamma$ is the gyromagnetic ratio, $\alpha$ is the Gilbert damping constant, and $M_{\rm s}$ is the saturation magnetization.
For the numerical simulations, we adopted a saturation magnetization of $M_{\rm s} = 4.8 \times 10^5 \, (1.4 \times 10^5)$ A/m, which corresponds to that of Ni\,(YIG).
The effective field acting on the $i$ cite $\mathbf{H}_{{\rm eff},i}$ is given by
\begin{equation}
\mathbf{H}_{{\rm eff},i} = \mathbf{H}_{{\rm ext},i}+ \mathbf{H}_{{\rm ani},i} + \mathbf{H}_{{\rm exc},i} + \mathbf{H}_{{\rm dd},i}
\end{equation}
where $\mathbf{H}_{{\rm ext},i}$ is the external input field, $\mathbf{H}_{{\rm ani},i}$ is the uniaxial anisotropy field, $\mathbf{H}_{{\rm exc},i}$ is the exchange field (coupling nearest neighbors), and $\mathbf{H}_{\rm {dd},i}$ is the dipole field.
The $\mathbf{H}_{{\rm dd},i}$ is given by
\begin{equation}
\mathbf{H}_{{\rm dd},i} = -\sum_{j} N_{ij} \mathbf{M}_j
\label{Hdd}
\end{equation}
where $\mathbf{M}_j$ is the magnetic moment of magnet $j$ and $N_{ij}$ is the demagnetization tensor between $i$ and $j$.
The total effective field on the magnet $i$ is the sum of all these contributions.
Micromagnetic simulations were performed using the Object Oriented MicroMagnetic Framework (OOMMF)\cite{OOMMF}.
We employ the 4th order Runge-Kutta method for calculating the time evolution with the time step $0.1$\,ps.
\begin{figure}[t]
\includegraphics[scale=0.4]{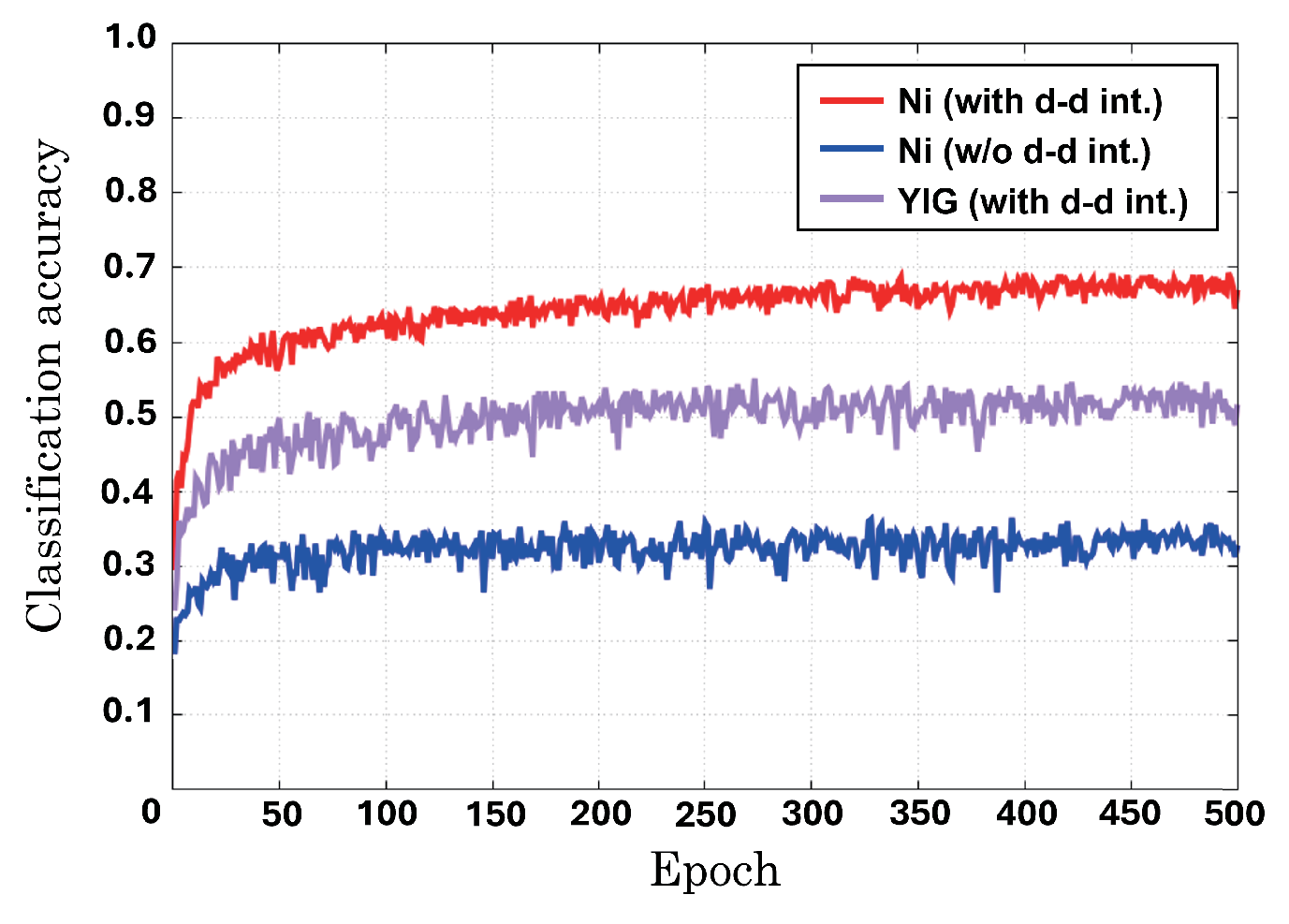}
\caption{
Classification accuracy for two magnetic materials (Ni and YIG) that have different saturation magnetization.
For the Ni impurities, results of $50\%$ impurities system with and w/o the dipole-dipole interaction are shown.
}
\label{fig2}
\end{figure}

We prepare the external magnetic field as an input hand-written data provided by MNIST\cite{MNIST}.
The gray scale image data of handwritten numbers $0$ to $9$ is encoded to the spatially dependent magnetic field as shown in the Fig. \ref{fig1}.
The maximum magnetic field is set to $1$\,T, and applied in the $x$ direction.
The magnetic field is applied during the first $1$\,ns and we record the magnetization dynamics after switching off the magnetic field $1\sim 2$\,ns for every time step $\Delta t = 0.01$\,ns.
In this work, we use the spatially averaged magnetization as the output data
\begin{equation}
\langle M^{\mu}(t) \rangle = \frac{1}{N_{\rm imp}} \sum_{i=1}^{N_{\rm imp}} M_{i}^{\mu}(t) \quad (\mu=x, y, z).
\end{equation}
Here, $N_{\rm imp}$ is the number of the magnetic impurities.
Although many physical RC systems uses the spatio-temporal output data for achieving the high accuracy of the classification, we will show that spatially averaged data include enough information for performing the classification.
Due to the random distribution of the magnetic impurities and the long-range dipole coupling, each magnetic moment precesses a different phase and frequency.
The spatial average does not cancel out these signals and it creates a complex interference pattern that contains the superposition of all individual non-linear responses.
The dipole interaction acts as a ``multiplexer" that encodes the high-dimensional spatial input into the temporal fluctuations of the average magnetization.
These averaged magnetization are linearly transformed by a trained output weight matrix ($W$) to produce the final output $(o_\ell)$
\begin{equation}
o_{\ell} = \sum_{\mu, n}\left< M^\mu(t_n) \right> W^{\mu}_{l,n},
\end{equation}
where $n$ represents the total time step.
The total time step is $100$ for the time evolution $1.0$\,ns with $\Delta t=0.01$\,ns.
We use the three components ($M^x, M^y, M^z$) of the magnetization, then $300$ input data are used for calculating one sample.
The predicted probability $y_{\ell}$ for class $\ell$ is obtained via the softmax function applied to the output $o_{\ell}$.

To train the weight matrix, we use the Categorical Cross-Entropy Loss
\begin{equation}
L = - \frac{1}{N} \sum_{k=1}^{N} \sum_{\ell=1}^{10} x_{\ell}^{(k)} \log(y_{\ell}^{(k)})
\end{equation}
where $N$ is the number of training samples, and $x_{\ell}^{(k)}$ is the target label expressed by one-hot vector for the $k$-th sample.
We used the Stochastic Gradient Descent (SGD) optimizer to update the output weights $W_{l,n}^\mu$.
The learning rate was set to 0.01.
We use $1600$ samples for training set  and $400$ samples for validation set.
Note that the reservoir (the magnetic impurity system) itself is not trained.


First, we show that the dipole-dipole interaction is quite useful for realizing the RC by magnetic systems.
We show the classification results for the system with and without the dipole-dipole interaction for $50\%$ of a magnetic impurity (Ni) density in Fig.~\ref{fig2}.
The reservoir with the dipole-dipole interaction achieved a stable classification accuracy of $70\%$ after $500$ training epochs using the training dataset.
This rapid convergence and high accuracy demonstrate the effective separation of temporal input patterns enabled by the complex magnetization dynamics.
Obtained accuracy is quite high compared with that without the dipole-dipole interaction $(\sim 35\%)$.
Without dipole coupling, the ferromagnetic interaction between neighboring magnetic impurities produces the nonlinear network that gives the lower accuracy.

We also compared the RC performance by using different magnetic material that has different saturation magnetization.
The large saturation magnetization induces the large dipole field represented by Eq.(\ref{Hdd}).
It is expected that the large saturation magnetization allows the long-range information propagation that makes the spin network complex enough to convert the spatial-domain input data into time-domain output data.
Figure ~\ref{fig2} shows that the large saturation magnetization (Ni) can achieve the high accuracy compared with that of the small saturation magnetization (YIG).

\begin{figure}[t]
\includegraphics[scale=0.38]{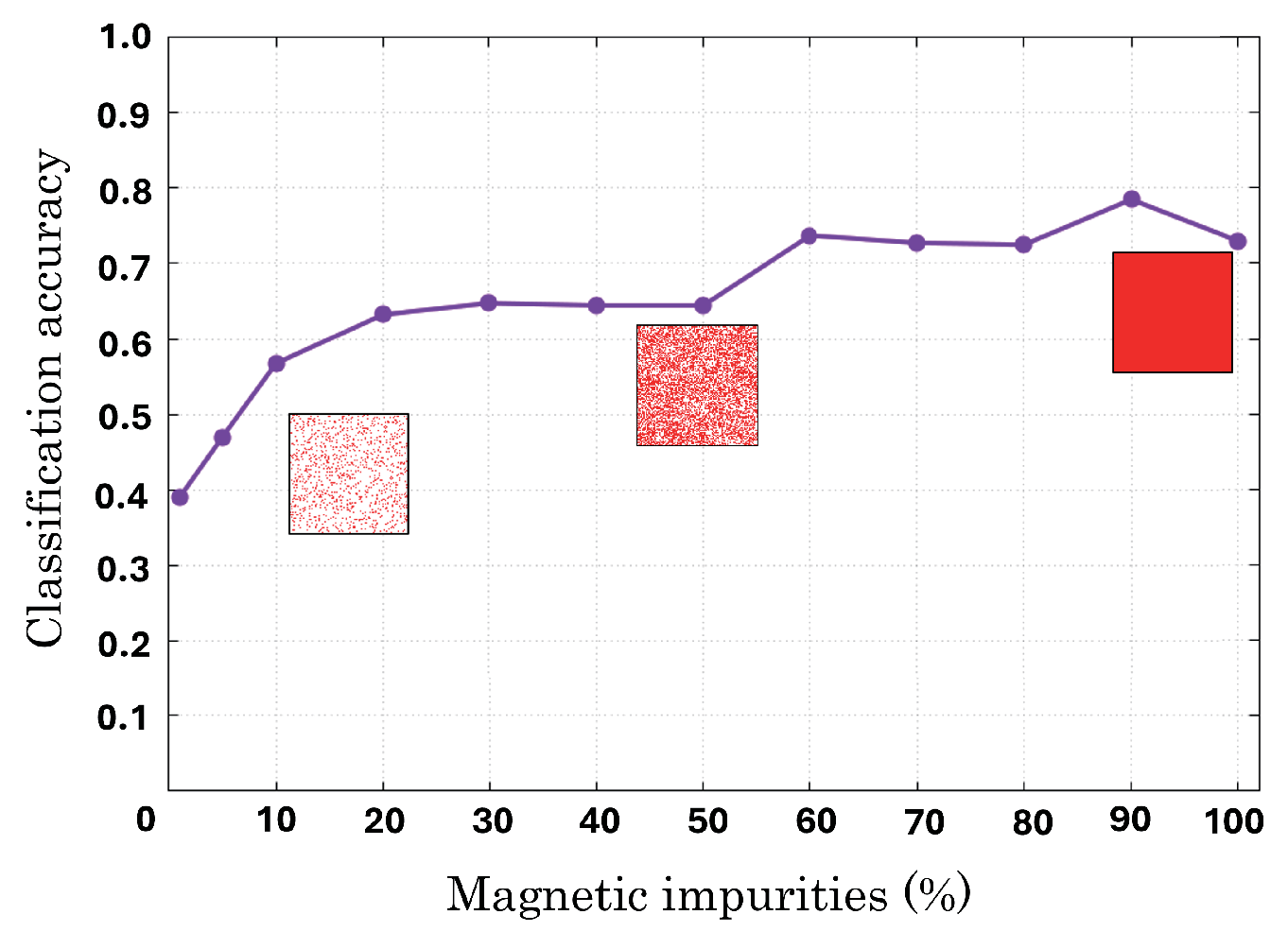}
\caption{
Classification accuracy for several impurities concentration (Ni).
}
\label{fig3}
\end{figure}

Our proposed system consists of magnetic impurities that does not require advanced microfabrication equipment.
We varied the concentration of magnetic impurities $c$ from 1$\%$ to 100$\%$ (continuous film).
Figure \ref{fig3} shows that the higher accuracy (83\%) is obtained by high concentration ($c=90\%$).
The accuracy is obtained after 500 training epochs.
The accuracy tends to decrease by decreasing the concentration.
For lower concentration, the impurities are too far apart for the dipole interaction to provide sufficient coupling, leading to a loss of memory.
We confirm that the lower accuracy ($\sim 0.1$) is obtained below $c=0.05\%$.
The optimal performance at $c=90\%$ suggests that a certain degree of structural disorder and moderate coupling density are necessary to maximize the computational capacity.

\begin{figure}[b]
\includegraphics[scale=0.38]{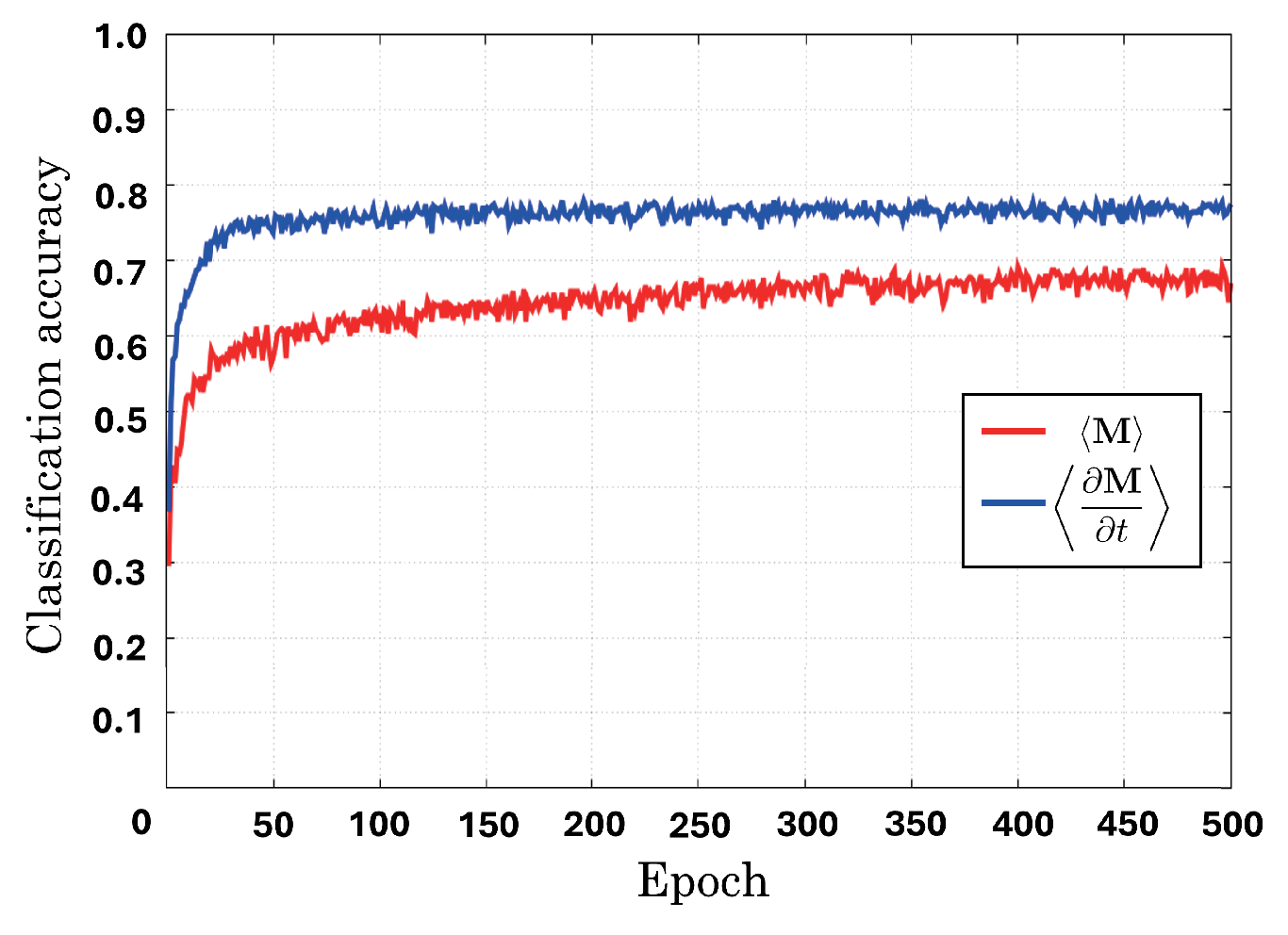}
\caption{
Classification accuracy for the Ni impurities (50\%).
Results of the magnetization component data and its time derivative data.}
\label{fig4}
\end{figure}


To increase the accuracy of the classification, one can employed the data manipulation for the leaning process.
We use the time derivative of the magnetization dynamics $ \frac{\partial \left<\mathbf{M} \right>}{\partial t}$ instead of the magnetization component $\left<\mathbf{M} \right>$.
The time derivative is calculated by the $(\left<\mathbf{M}(t_{n+1}) \right> - \left<\mathbf{M}(t_{n}) \right>)/\Delta t$.
The time step $(\Delta t = t_{n+1} - t_n)$ is $0.01$\,ns.
Figure \ref{fig4} shows that the result by using the $ \frac{\partial \left<\mathbf{M} \right>}{\partial t}$ and $\left<\mathbf{M} \right>$.
In these calculation, the number of the input data (the total time step) for one sample is fixed to $300$.
One can see that $75\%$ of accuracy is obtained by using the data of $ \frac{\partial \left<\mathbf{M} \right>}{\partial t}$.
The time derivative is calculated by the data of $\left<\mathbf{M} \right>$.
The results suggest that the data manipulation has an advantage.

We have investigated the physical reservoir computing (RC) by using the magnetization dynamics of magnetic impurities coupled by the dipole-dipole interaction.
The proposed system is easy to fabricate and the measurable data is used for the classification of the hand-written digits.
We simulated the magnetization dynamics, from which we recognize the digits 0 to 9.
The results indicate that the input spatial-domain data $H_x(x, y)$ is converted to the time-domain output data $\left<\mathbf{M}(t) \right>$ through the non-linear magnetization dynamics.
The strength of the dipole-dipole coupling is determined by the value of the saturation magnetization, and the higher saturation magnetization provides higher accuracy.
Our proposed system paves the way to achieve the spintronics RC device without nanofabrication.

\begin{acknowledgments}
The authors are grateful to T. Kawarabayashi for valuable discussions.
This work was supported by JSPS KAKENHI Grant Numbers 21H05016, 22H05114 and 23K21077.
\end{acknowledgments}


\end{document}